\documentclass[aps,pre,reprint,superscriptaddress,amsmath,amssymb,preprintnumbers]{revtex4-2}

\usepackage{graphics,graphicx}
\begin{document}
\preprint{Physical Review E}

\title{Condition for \boldmath{$1/f$} noise to occur along with an example for a diffusion equation}

\author{Hideaki Mouri}
\affiliation{Meteorological Research Institute, Nagamine, Tsukuba 305-0052, Japan}
\affiliation{Tsukuba University of Technology,  Amakubo, Tsukuba 305-8520, Japan}

\begin{abstract}
While $1/f$ noise is ubiquitous and has been found in various systems, its physics remains uncertain. From an analytical study of an ordinary diffusion equation, we find an additional example of the $1/f$ noise. The formula for this example, together with existing knowledge about scaling in fluid turbulence, implies a necessary and sufficient condition for the occurrence of any stationary $1/f$ noise. That is, the noise needs to be characterized by two constant frequencies of $f_{\rm low} \ll f_{\rm high}$. For a frequency range from $f = f_{\rm low}$ to $f_{\rm high}$, it is further needed that, except for the mean amplitude of the noise, there is no other constant parameter. Then, at $f_{\rm low} \ll f \ll f_{\rm high}$, the noise scales asymptotically as $1/f$. Being statistical and simple, our condition applies to any system~and~hence~explains the ubiquity of the $1/f$ noise. It is also applicable to some systems with noise of $\alpha \ne 1.0$ for $1/f^{\alpha}$\!,~via intermittency analogous to that of the turbulence.
\end{abstract}

\maketitle

\section{Introduction} \label{S1}

One hundred years ago, for the output voltage of a vacuum tube, Johnson \cite{j25} found a phenomenon now known as $1/f$ noise. This voltage fluctuated randomly with an intensity inversely proportional to the frequency $f$, as long as $f$ was low enough. Since then, the same noise has been found in various materials, e.g., metals and semiconductors, used in electrical devices such as resistors and diodes \cite{p78, dh81, w88, pgfa14}.

There are other examples of $1/f$ noise \cite{p78}. It also occurs in the sand flow of an hourglass \cite{sv74}, traffic on an~expressway \cite{mh76}, the flow of the River Nile \cite{mw69}, ocean currents in the Pacific \cite{thwb74}, electric currents in the ionosphere of the Earth \cite{tsiggas90}, the solar wind \cite{mg86,b06}, accretion onto a black hole \cite{ngmprd81, lwpe87}, and so on. Thus, albeit absent in the voltages of some materials \cite{w88} and the flows of some rivers~\cite{mw69}, $1/f$ noise is ubiquitous.

Despite many years of study \cite{p78, dh81, w88, pgfa14, b37, ba66, h68, m80, ms82, btw87, krga06, mn23}, the physics of $1/f$ noise remains uncertain. It is unlikely to be universal \cite{dh81, w88, pgfa14}, given the diverse variety of the above sources of the $1/f$ noise. Nevertheless, for the underlying statistics, universality is expected as discussed here.

We consider that $1/f$ noise is stationary. As for nonstationarity, there is no convincing observation \cite{dh81, w88, pgfa14}.~The stationary $1/f$ noise requires a low-frequency cutoff so as not to diverge in the limit $f \rightarrow 0$. Such a cutoff has been found in some systems \cite{sv74, ngmprd81, mg86}, but not yet in electrical devices \cite{j25, p78, dh81, w88, pgfa14}. They exhibit $1/f$ noise over several orders of magnitude in frequency, implying that the frequency of the cutoff is very low.

The stationary and power-law scaling like that of $1/f$ noise is not rare, especially in fluids, dating back one hundred years to Richardson's law for the turbulent dispersion \cite{r26}. By using the knowledge accumulated there~\cite{r26, l32, k41, ll59, b59, k62, t76, pa77, l78, agha84, sd98, lx10, mmym17, mm19, isy17}, we would understand more about $1/f$ noise.

First, in fluid turbulence, the exponent of a power law is sensitive to the physics or even to the statistics \cite{r26, k41, ll59, b59, k62, pa77, agha84, sd98, lx10, isy17}. Although $1/f$ noise is often defined as $\alpha \simeq 0.8$--$1.4$ for $1/f^{\alpha}$ \cite{w88}, this range is too wide to expect any universality. Indeed, in fluid turbulence, buoyancy induces temperature fluctuations of $\alpha = 7/5 = 1.4$ under optimal conditions \cite{b59, lx10}, but not of $\alpha = 1.0$. To avoid this and other similar cases, we focus on the standard~value~$\alpha = 1.0$. It is still observed in a variety of systems \cite{j25, p78, dh81, w88, pgfa14, sv74, mh76, mw69, thwb74, tsiggas90, mg86, b06, ngmprd81, lwpe87}.

The exponent $\alpha = 1.0$ imposes a high-frequency cutoff onto the power law, beyond which the noise decays faster, since its intensity integrated over all frequencies needs to be finite without diverging as $\ln f$ \cite{dh81, w88}. This is actually observed when the observation frequency is high enough and the background is low enough  \cite{sv74, tsiggas90, b06}.

Second, for a random process $j(t)$ with a mean amplitude $j_{\rm c}$, if $j(t)/j_{\rm c}$ is characterized by two constant frequencies of $f_{\rm low} \ll f_{\rm high}$ and if no other constant exists at $f = f_{\rm low}$ to $f_{\rm high}$, it follows from a turbulence theory \cite{pa77} that the spectral intensity $I_j$, to be defined in Sec.~\ref{S2}, is written as $\varpropto j_{\rm c}^2/f$ at $f_{\rm low} \ll f \ll f_{\rm high}$. Since $f_{\rm low}$ and $f_{\rm high}$ also correspond to the above two cutoffs, we expect that the presence of such frequencies is necessary and sufficient for $1/f$ noise in general to occur. This condition is statistical and hence simple, while only simplicity would explain the ubiquity of the $1/f$ noise \cite{mn23}.

To confirm our expectation, we study a diffusion equation in Sec.~\ref{S3}. For a stationary case, it is found analytically that the diffusion flux $j(t)$ does exhibit $1/f$ noise in a manner expected above. Then, in Sec.~\ref{S4}, we discuss a general yet rigorous formulation of our expectation. The result is applied to one of the basic models of $1/f$ noise \cite{p78, dh81, w88, pgfa14}, i.e., a random superposition of random pulses \cite{b37, h68}. Finally, in Sec.~\ref{S5}, we conclude with a remark that the same is expected for some cases of $\alpha \ne 1.0$, at least as an approximation, if we consider intermittency~as of the fluid turbulence \cite{k62, agha84, sd98, isy17}.

\section{Definitions} \label{S2}

Consider real-valued, stationary, and zero-mean fluctuations. For those of a random process $\varphi (t)$, where $t$ is the time, the Fourier transformation is defined as
\begin{subequations} \label{eq1}
\begin{equation} \label{eq1a}
\tilde{\varphi}(\omega) = \! \int^{+\infty}_{-\infty}  \!\! \varphi(t) \exp (-i \omega t) dt .
\end{equation}
We have adopted the angular frequency $\omega = 2\pi f$. The inverse transformation is
\begin{equation} \label{eq1b}
{\varphi}(t) = \! \int^{+\infty}_{-\infty}  \!\! \tilde{\varphi}(\omega) \exp (i \omega t) \frac{d \omega}{2\pi}.
\end{equation}
\end{subequations}
Since $\varphi (t)$ is real-valued, we have $\tilde{\varphi}^{\ast}(\omega) = \tilde{\varphi}(-\omega)$. The spectral intensity $I_{\varphi}(\omega)$ is then defined for $\omega > 0$ as
\begin{equation} \label{eq2}
I_{\varphi}(\omega) \delta (\omega - \omega' ) 
=\! \frac{\langle \tilde{\varphi}(\omega)\tilde{\varphi}^{\ast}(\omega') \rangle\! + \!\langle \tilde{\varphi}^{\ast}(\omega)\tilde{\varphi}(\omega') \rangle}{2\pi} .
\end{equation}
While $\delta (\omega) = (1/2\pi ) \int^{+\infty}_{-\infty} \exp (-i\omega t) dt$ is the Delta function, $\langle ... \rangle$ denotes an ensemble average.

These definitions are used to write the convolution of two processes $\varphi(t)$ and $\psi(t)$,
\begin{equation} \label{eq3}
\int^{+\infty}_{-\infty}  \!\! \varphi(t-t') \psi(t') dt' = \! \int^{+\infty}_{-\infty} \! \!\! \tilde{\varphi}(\omega) \tilde{\psi}(\omega) \exp (i \omega t) \frac{d \omega}{2\pi} .
\end{equation}
Likewise, from $\langle \varphi \rangle = 0$, the Wiener-Khinchin theorem is
\begin{equation} \label{eq4}
I_{\varphi}(\omega) = 4\! \int^{+\infty}_0 \!\! \langle \varphi(t+\tau) \varphi(t) \rangle \cos (\omega \tau) d\tau.
\end{equation}
The two-time correlation $\langle \varphi(t+\tau) \varphi(t) \rangle$ is independent of $t$ and of the sign for $\tau$, i.e., $ \langle \varphi(t+\tau)\varphi(t) \rangle = \langle \varphi(t-\tau)\varphi(t) \rangle$, because $\varphi(t)$ is stationary. We use $[ \int^{\infty}_0\! I_{\varphi}(\omega) d \omega ]^{1/2}$ as the mean amplitude of $\varphi (t)$ in Sec.~\ref{S4}.

\section{Analytical Example} \label{S3}

\begin{figure}[tbp]
\begin{center}
\resizebox{8.55cm}{!}{\includegraphics*[3.9cm,20.0cm][17.0cm,26.0cm]{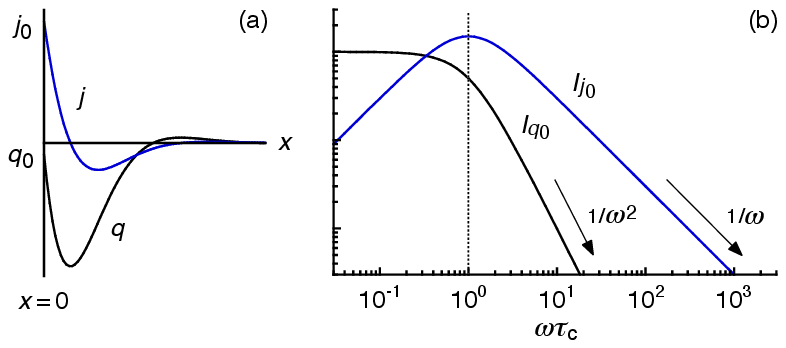}}
\caption{\label{f1} (a) Examples of Fourier modes in Eq.~(\ref{eq10}) at a certain time $t$ against the position $x$. (b) Spectral intensities $I_{q_0}$~and $I_{j_0}$ in Eqs.~(\ref{eq7}) and (\ref{eq11}) against the normalized frequency $\omega \tau_{\rm c}$.}
\end{center}
\end{figure}

Figure \ref{f1}(a) shows our setup for the diffusion of a quantity $q$ from a boundary at $x = 0$ into a medium at $x > 0$. We write a one-dimensional diffusion equation as
\begin{subequations} \label{eq5}
\begin{equation} \label{eq5a}
\frac{\partial}{\partial t}q(x, t) = D \frac{\partial^2}{\partial x^2}q(x, t) ,
\end{equation}
with the diffusion flux
\begin{equation} \label{eq5b}
j(x, t) = -D \frac{\partial}{\partial x}q(x, t) .
\end{equation}
\end{subequations}
The diffusion coefficient $D$ is a constant with a dimension {\sf L}$^{\sf 2}$\!{\sf /T}, where {\sf L} and {\sf T} denote the length and the time. This equation is ordinary and is distinct from those in previous models of anomalous diffusion to obtain $1/f$ noise for the quantity $q$ \cite{dh81}.

Under a stationary condition where $q$ at $x = 0$ fluctuates in a random manner, we study the flux $j$ at the~same boundary $x = 0$. These are written as
\begin{equation} \label{eq6}
q_0(t) = q(x, t) \vert_{x = 0} 
\ \ \mbox{and} \ \
j_0(t) = j(x, t) \vert_{x = 0}.
\end{equation}
To specify the fluctuations of $q_0(t)$, we adopt a correlation function that decays exponentially with a timescale $\tau_{\rm c}$,
\begin{subequations} \label{eq7}
\begin{equation} \label{eq7a}
\langle q_0(t+\tau) q_0(t) \rangle = \langle q_0^2 \rangle \exp (-\vert \tau \vert / \tau_{\rm c} ) .
\end{equation}
This correlation leads to a Lorentzian spectrum via Eq. (\ref{eq4}),
\begin{equation} \label{eq7b}
I_{q_0}(\omega) = \frac{4\tau_{\rm c}}{1+ (\omega \tau_{\rm c})^2}  \langle q_0^2 \rangle .
\end{equation}
\end{subequations}
Together with $D$ in Eq.~(\ref{eq5}), two constants $\tau_{\rm c}$ and $\langle q_0^2 \rangle$ in Eq.~(\ref{eq7}) are to make up another constant $j_{\rm c}\! =\! \sqrt{D \langle q_0^2 \rangle /\tau_{\rm c}}$ for the $1/f$ noise $\varpropto j_{\rm c}^2/\omega$ of the flux $j_0(t)$ [Eq.~(\ref{eq11b})].

The formula of Eq.~(\ref{eq5}) has various applications. Not only to mass transfer in a fluid \cite{ll59} and heat transfer in a fluid \cite{ll59} or in a solid, it applies to momentum transfer in a flow over an oscillating plane \cite{ll59, l32} or within a thin tube \cite{l32}, penetration of electromagnetic waves into a conducting medium \cite{pp62}, transfer of magnetic field lines across such a medium \cite{pp62}, and so on. These are likely to include unknown cases of $1/f$ noise. Their equivalent electrical circuit is described in Appendix. Furthermore, Eq.~(\ref{eq5}) might apply to some of the known cases, particularly those for flows \cite{mh76, mw69, thwb74, tsiggas90, mg86, b06, lwpe87}.

\subsection{Behavior in the frequency domain} \label{S3a}

To solve Eq.~(\ref{eq5}), we write $q(x, t)$ as a superposition of Fourier modes \cite{ll59, l32},
\begin{equation} \label{eq8}
q(x, t) = \! \int^{+\infty}_{-\infty}  \!\! \tilde{q}_0 (\omega) \exp(ikx + i \omega t) \frac{d \omega}{2\pi}.
\end{equation}
The wavenumber $k$ is dependent on the angular frequency $\omega$. By substituting Eq.~(\ref{eq8}) into Eq.~(\ref{eq5a}),
\begin{equation} \label{eq9}
i\omega = -Dk^2
\ \ \mbox{or} \ \
k = \pm \sqrt{\frac{\omega}{iD}} .
\end{equation}
Because of $\sqrt{i} = (1+i)/\sqrt{2}$ and hence $ik = \pm i \sqrt{\omega/iD} = \pm \sqrt{i\omega/D} = \pm(1+i) \sqrt{\omega/2D}$, the minus sign is adopted~so that $\exp (ikx)$ does not increase with the distance $x$ from the boundary. We substitute Eq.~(\ref{eq9}) into Eq.~(\ref{eq8}),
\begin{subequations} \label{eq10}
\begin{equation} \label{eq10a}
q(x, t) = \! \int^{+\infty}_{-\infty} \!\!\! \tilde{q}_0 (\omega) \exp\! \left(\! -\sqrt{i\omega/D}\, x + i \omega t \right)\! \frac{d \omega}{2\pi} .
\end{equation}
The modes do decay within single wavelengths [Fig.~\ref{f1}(a)]. By substituting Eq.~(\ref{eq10a}) into Eq.~(\ref{eq5b}),
\begin{equation}  \label{eq10b}
j(x, t) = \! \int^{+\infty}_{-\infty} \!\!\! \tilde{q}_0 (\omega) \sqrt{i\omega D} \exp\! \left(\! -\sqrt{i\omega/D}\, x + i \omega t \right)\! \frac{d \omega}{2\pi} .
\end{equation}
\end{subequations}
The spatial differentiation in Eq.~(\ref{eq5b}) has multiplied each of the Fourier modes by a factor $\varpropto \sqrt{\omega}$. It corresponds to a factor $\varpropto \omega$ in the spectral intensity.

Through Eq.~(\ref{eq10b}) at the boundary $x = 0$, the spectral intensity $I_{q_0}(\omega)$ in Eq.~(\ref{eq7b}) leads to that of $j_0(t)$,
\begin{subequations} \label{eq11}
\begin{equation} \label{eq11a}
I_{j_0}(\omega) = \frac{4\omega \tau_{\rm c}}{1+ (\omega \tau_{\rm c})^2} D\langle q_0^2 \rangle .
\end{equation}
Figure \ref{f1}(b) compares $I_{q_0}\!(\omega)$ with $I_{j_0}\!(\omega)$. As $\omega \rightarrow +\infty$, we observe $I_{q_0} \varpropto 1/\omega^2$ and $I_{j_0} \varpropto 1/\omega$. The latter is written as
\begin{equation} \label{eq11b}
I_{j_0}(\omega) = \frac{4D\langle q_0^2 \rangle}{\omega \tau_{\rm c}}
\ \ \mbox{at} \
\omega \gg \frac{1}{ \tau_{\rm c} }.
\end{equation}
\end{subequations}
Thus, the diffusion flux $j_0(t)$ at the boundary $x = 0$~fluctuates as $1/f = 2\pi/\omega$.

Since $I_{j_0}(\omega)$ in Eq.~(\ref{eq11}) does not have a high-frequency cutoff, $\int^{\infty}_0\! I_{j_0}(\omega) d\omega$ is divergent. Nevertheless, in practice \cite{w88, pp62}, a limitation on the model for the diffusion equation [Eq.~(\ref{eq5})] or for the boundary condition [Eq.~(\ref{eq7})] imposes the cutoff at, say, $\omega = \omega_{\rm high}$. We regard Eq.~(\ref{eq11}) as a result of already taking the limit $\omega_{\rm high} \rightarrow +\infty$ (see Appendix for a case of such a limit).

This $1/f$ noise of the flux $j_0(t)$ confirms our expectation in Sec.~\ref{S1}. If the dimension of $q_0(t)$ is {\sf Q}, that of $j_0(t)$ is {\sf QL/T} [Eq.~(\ref{eq5b})]. As for its amplitude $j_{\rm c}$ made up of the given constants $D$ in {\sf L}$^{\sf 2}$\!{\sf /T}, $\tau_{\rm c}$ in {\sf T}, and $\langle q_0^2 \rangle$ in {\sf Q}$^{\sf 2}$\!,~we only have $\varpropto\! \sqrt{D \langle q_0^2 \rangle /\tau_{\rm c}}$. The scaling $\varpropto j_{\rm c}^2/\omega$ is realized at $\omega \gg \omega_{\rm low} = 1/\tau_{\rm c}$ [Eq.~(\ref{eq11b})]. Expectedly, in this~range, no constant parameter exists for the shape~of~the spectral intensity $I_{j_0}(\omega)$.

\subsection{Behavior in the time domain} \label{S3b}

The diffusion flux $j_0(t)$ is studied still more in the time domain \cite{ll59, ch62}, by rewriting Eq.~(\ref{eq10b}) at the boundary $x = 0$. Via the identity $\int^{+\infty}_0\! \exp(-i\omega t)dt/\sqrt{\pi t} = 1/\sqrt{i\omega}$, we use the convolution of Eq.~(\ref{eq3}) for $\tilde{\varphi}(\omega) = i\omega \tilde{q}_0(\omega)$ and $\tilde{\psi}(\omega) = 1/\sqrt{i\omega}$ or for $\varphi (t-t') = dq_0(t-t')/d(t-t')$ and $\psi (t') = 1/\sqrt{\pi t'}$ at $t' > 0$ with $\psi (t') = 0$ at $t' \le 0$,
\begin{subequations} \label{eq12}
\begin{equation} \label{eq12a}
j_0(t) = \!\! \int^{+\infty}_{0} \!\! \frac{dq_0(t-t')}{d(t-t')} \frac{dt'}{\sqrt{\pi t' /D}} 
       = \! \!\int^{t}_{-\infty} \!\!\!\!\! \frac{dq_0(t')}{dt'} \frac{dt'}{\sqrt{\pi (t-t')/D}} .
\end{equation}
For the fluctuations of $dq_0(t')/dt'$, we adopt the Ornstein-Uhlenbeck process,
\begin{equation} \label{eq12b}
\frac{dq_0(t')}{dt'} = - \frac{q_0(t')}{\tau_{\rm c}} + \sqrt{\frac{4 \langle q_0^2 \rangle}{\tau_{\rm c}}} w(t') .
\end{equation}
The unit white noise $w(t')$ has a dimension {\sf T}$^{\sf -1/2}$,
\begin{equation} \label{eq12c}
\langle w(t'+\tau) w(t') \rangle = \frac{\delta(\tau)}{2}
\ \ \mbox{so that} \ \
I_w(\omega) = 1 .
\end{equation}
\end{subequations}
Through $\tilde{q}_0(\omega) = \sqrt{4\langle q_0^2 \rangle /\tau_{\rm c}} \tilde{w}(\omega)/(i\omega+1/\tau_{\rm c})$, as derived from Eq.~(\ref{eq12b}), $I_w(\omega)$ in Eq.~(\ref{eq12c}) reproduces $I_{q_0}(\omega)$ in Eq.~(\ref{eq7b}).

In addition to the diffusion, Eq.~(\ref{eq12a}) serves as a solution for sound pressure from a line source \cite{l78}. The $1/f$ noise is expected also there.

The formula for $j_0(t)$ of Eq.~(\ref{eq12a}) is just the half-order differentiation of $q_0(t)$ \cite{l78, ch62}. Since $q_0(t)$ is obtained by integrating $w(t)$ in Eq.~(\ref{eq12b}), the flux $j_0(t)$ is explainable as the half-order integration of the white noise $w(t)$. It multiplies $\tilde{w}(\omega)$ by a factor $\varpropto 1/\sqrt{\omega}$, like a factor $\varpropto 1/\omega$ in a usual first-order integration. Thus, from $I_w(\omega) = 1$, we obtain $I_{j_0}(\omega)\! \varpropto\! 1/\omega\! =\! 1/2\pi f$. Such a calculus has~been used in some of previous models of $1/f$ noise \cite{p78, ba66}.

\begin{figure}[tbp]
\begin{center}
\resizebox{8.55cm}{!}{\includegraphics*[3.9cm,20.0cm][17.0cm,26.0cm]{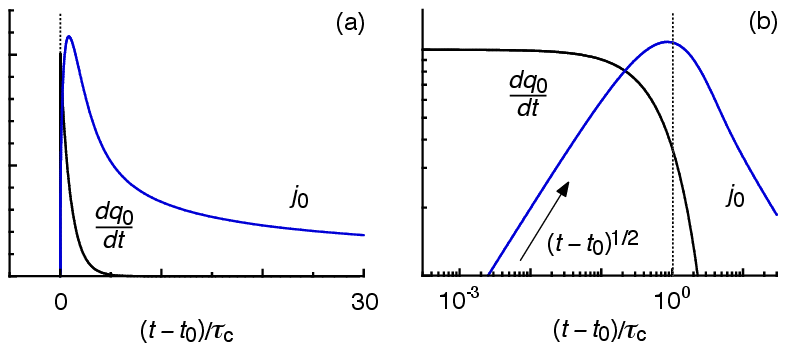}}
\caption{\label{f2} (a) Linear plot of $j_0$ and $dq_0/dt$ in Eq.~(\ref{eq13}), induced by an impulse at time $t_0$, against the normalized lapse time $(t-t_0)/\tau_{\rm c}$. (b) Same as in (a) but on logarithmic scales.}
\end{center}
\end{figure} 

Furthermore, Eq.~(\ref{eq12a}) describes the accumulation of responses $\varpropto 1/\sqrt{t-t'}$ to the individual fluctuations of $dq_0(t')/dt'$. The latter are due to the onset and relaxation of the individual impulses of the white noise $w(t'')$ at~$t'' < t' < t$ [Eq.~(\ref{eq12b})]. For an impulse at $t'' = t_0$, the response is
\begin{subequations} \label{eq13}
\begin{equation}
j_0(t_0 | t) = \!\!\int^{t}_{t_0} \! \frac{dq_0(t_0|t')}{dt'}  \frac{dt'}{\sqrt{\pi (t-t')/D}} .
\end{equation}
The relaxation of $q_0(t_0)\! = \!\sqrt{4 \langle q_0^2 \rangle / \tau_{\rm c}} w(t_0)dt_0$ is written as
\begin{equation}
\frac{dq_0(t_0|t')}{dt'}= - \frac{q_0(t_0)}{\tau_{\rm c}} \exp\! \left(\! -\frac{t'-t_0}{\tau_{\rm c}} \!\right) .
\end{equation}
\end{subequations}
Figure \ref{f2} shows the linear and logarithmic plots of $j_0(t_0 | t)$ and $dq_0(t_0|t)/dt$. The response $j_0(t_0 | t)$ is slow \cite{p78}. Especially at $t - t_0 \ll \tau_{\rm c}$ on the logarithmic plot, corresponding to the range $\omega \gg 1/\tau_{\rm c}$ of the $1/f$ noise, although $dq_0(t_0|t)/dt$ has already increased, $j_0(t_0 | t)$ is still increasing as $\int^t_{t_0} dt'/\sqrt{t-t'}\! \varpropto\! \sqrt{t-t_0}$. This slow response has transformed the white noise $w(t)$ into the $1/f$ noise of~the diffusion flux $j_0(t)$.

\section{Discussion} \label{S4}
Having confirmed the condition expected for the occurrence of $1/f$ noise with an example for a diffusion~equation, we now discuss its general formulation. The~result~is applied to another example of the noise model.

\vspace{-1mm}

\subsection{General formulation} \label{S4a}

Consider a stationary noise process $j(t)$ with two constant timescales of $\tau_{\rm small} \ll \tau_{\rm large}$. The mean amplitude $j_{\rm c}$ is finite and is given via $j_{\rm c}^2 = \int^{\infty}_0\! I_j(\omega) d \omega$. No other constant parameter exists at least at $\omega = 1/ \tau_{\rm large}$ to $1/ \tau_{\rm small}$. Then, in a range of $1/\tau_{\rm large} \ll \omega \ll 1/\tau_{\rm small}$,~the noise scales asymptotically as $1/f = 2\pi/\omega$. Its spectral intensity is
\begin{equation} \label{eq14}
I_j(\omega) = C \frac{j_{\rm c}^2}{\omega}
\ \ \mbox{at} \ \
\frac{1}{\tau_{\rm large}} \ll \omega \ll \frac{1}{\tau_{\rm small}} .
\end{equation}
The constant $C$ is dimensionless. In the limit $\tau_{\rm small} \rightarrow 0$, Eq.~(\ref{eq14}) corresponds to Eq.~(\ref{eq11b}) via $\tau_{\rm large} = \tau_{\rm c}$ and~$j_{\rm c} = \sqrt{D \langle q_0^2 \rangle /\tau_{\rm c}}$ with $C = 4$.

\begin{figure}[tbp]
\begin{center}
\resizebox{8.55cm}{!}{\includegraphics*[3.9cm,20.0cm][17.0cm,26.0cm]{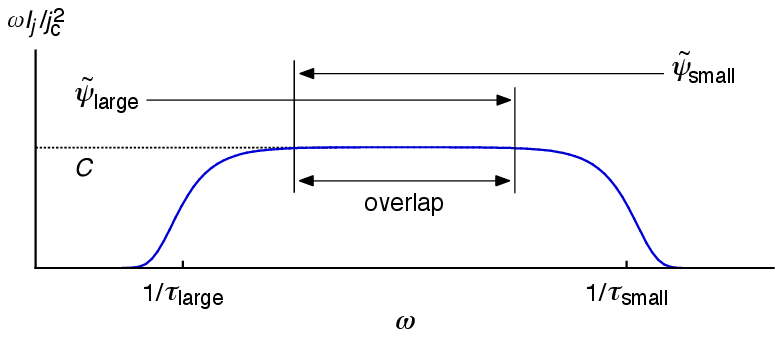}}
\caption{\label{f3} Overlap between $\tilde{\psi}_{\rm large}$ and $\tilde{\psi}_{\rm small}$, i.e., two solutions for $\omega I_j/j_{\rm c}^2$ in Eq.~(\ref{eq15}) at the angular frequencies of $1/\tau_{\rm large} \lesssim \omega \ll 1/\tau_{\rm small}$ and $1/\tau_{\rm large} \ll \omega \lesssim 1/\tau_{\rm small}$.}
\end{center}
\end{figure}

The derivation of Eq.~(\ref{eq14}) is based on the theory for a power law $\varpropto 1/k$ in fluid turbulence by Perry and Abell \cite{pa77}. We normalize $I_j(\omega)$ as $\omega I_j(\omega)/j_{\rm c}^2$. Such a dimensionless function needs to be specified by dimensionless parameters alone. Those available here are~$\omega\tau_{\rm large}$~at~$\omega \simeq 1/\tau_{\rm large}$ and $\omega\tau_{\rm small}$ at $\omega \simeq 1/\tau_{\rm small}$,
\begin{subequations} \label{eq15}
\begin{equation} \label{eq15a}
{\everymath{\displaystyle}
\frac{\omega I_j(\omega)}{j_{\rm c}^2} 
= \left\{ \begin{array}{ll}
                      \tilde{\psi}_{\rm large}(\omega\tau_{\rm large})& \mbox{at}\ \frac{1}{\tau_{\rm large}} \lesssim \omega \ll \frac{1}{\tau_{\rm small}} ,  \\
   \rule{0ex}{5.5ex}  
                      \tilde{\psi}_{\rm small}(\omega\tau_{\rm small})& \mbox{at}\ \frac{1}{\tau_{\rm large}} \ll \omega \lesssim \frac{1}{\tau_{\rm small}} .
          \end{array}
  \right.
}
\end{equation}
As shown for a range of $1/\tau_{\rm large} \ll \omega \ll 1/\tau_{\rm small}$ in Fig.~\ref{f3}, the two functions $\tilde{\psi}_{\rm large}(\omega\tau_{\rm large})$ and $\tilde{\psi}_{\rm small}(\omega\tau_{\rm small})$ overlap with each other. They become a constant $C$ independently of $\omega\tau_{\rm large}$ and $\omega\tau_{\rm small}$,
\begin{align} \label{eq15b}
\tilde{\psi}_{\rm large}(\omega\tau_{\rm large}) = \
& \tilde{\psi}_{\rm small}(\omega\tau_{\rm small}) = C \nonumber \\
&\qquad \mbox{at} \
\frac{1}{\tau_{\rm large}} \ll \omega \ll \frac{1}{\tau_{\rm small}} .
\end{align}
\end{subequations}
Since Eq.~(\ref{eq15b}) is equivalent to Eq.~(\ref{eq14}), the presence of $\tau_{\rm small}$ and $\tau_{\rm large}$, along with the absence of any other~constant, is sufficient for $j(t)/j_{\rm c}$ to exhibit $1/f$ noise.

The timescales $\tau_{\rm small}$ and $\tau_{\rm large}$ correspond to the frequencies of cutoffs of the $1/f$ noise, i.e., $\omega_{\rm low} = 2 \pi f_{\rm low} = 1/\tau_{\rm large}$ and $\omega_{\rm high}\! =\! 2 \pi f_{\rm high}\! =\! 1/\tau_{\rm small}$. As has been noted in Sec.~\ref{S1}, such cutoffs are also necessary for $\int^{\infty}_0\! I_j(\omega) d \omega$ and $I_j(\omega)$ so as not to diverge to infinity.

This necessary and sufficient condition applies to any system, irrespective of the physics, so that its simplicity explains why $1/f$ noise is ubiquitous \cite{j25, p78, dh81, w88, pgfa14, sv74, mh76, mw69, thwb74, tsiggas90, mg86, b06, ngmprd81, lwpe87}. If it does not hold, e.g., due to an insufficient separation between $\tau_{\rm small}$ and $\tau_{\rm large}$ or a presence of another constant parameter, $1/f$ noise would not occur.

\subsection{Application to the random pulse model} \label{S4b}

Particularly for electrical devices \cite{p78, dh81, w88, pgfa14}, $1/f$ noise is often modeled as a random superposition of pulses that are randomly stretched in time \cite{b37, h68}. Their shapes remain the same, e.g., exponential decay. If the stretching factor is distributed over a wide range in an optimal manner, the model reproduces the $1/f$ noise.

This model still has room for a generalization. Under the condition of Sec.~\ref{S4a}, it is not necessary to assume the same pulse shape or to optimize the distribution of the stretching factor or of the pulse width.

On the basis of a model of turbulence eddies \cite{t76, mmym17, mm19}, we use a mean noise amplitude $j_{\rm c}$ and dimensionless functions $\varphi_{\rm pulse}$ to describe a superposition of pulses as
\begin{equation} \label{eq16}
j(t)= j_{\rm c} \!\! \sum_{s_0, t_0, \tau_0} \!\! \varphi_{\rm pulse}^{(s_0)}\!\left(\frac{t-t_0}{\tau_0}\right)\!
\ \ \mbox{for} \ 
\tau_{\rm small} < \tau_0 \le \tau_{\rm large} .
\end{equation}
The timescales $\tau_{\rm small}$ and $\tau_{\rm large}$ are identical to those used in Sec.~\ref{S4a}. Later, we increase their separation. There~is no other constant parameter.

The pulse onset time $t_0$ and the pulse width $\tau_0$ are random. As for the pulse shape, we have used a dimensionless random parameter $s_0$ with $\varphi_{\rm pulse}^{(-s_0)} = -\varphi_{\rm pulse}^{(+s_0)}$.~These three are independent of one another.

For fluctuations due to pulses of a given width $\tau_0$, with $\sum_{s_0} \langle \varphi_{\rm pulse}^{(s_0)} \rangle = \sum_{s_0 > 0} \langle \varphi_{\rm pulse}^{(s_0)} \rangle + \sum_{s_0 < 0} \langle \varphi_{\rm pulse}^{(s_0)} \rangle = 0$, the two-time correlation is written in a dimensionless form as
\begin{subequations} \label{eq17}
\begin{equation} \label{eq17a}
\sum_{s_0} \left\langle\! \varphi_{\rm pulse}^{(s_0)}\! \left(\frac{t-t_0+\tau}{\tau_0}\right) 
                      \varphi_{\rm pulse}^{(s_0)}\! \left(\frac{t-t_0}     {\tau_0}\right)\! \right\rangle 
=
\phi_{\rm pulse}\! \left(\frac{\tau}{\tau_0}\right)\!.
\end{equation}
This correlation no longer depends on the shape $s_0$ or on the time $t_0$ and is a function of $\tau/\tau_0$ alone. We substitute Eq.~(\ref{eq17a}) into Eq.~(\ref{eq4}) to obtain the spectral intensity,
\begin{equation} \label{eq17b}
4 \!\! \int^{+\infty}_0\hspace{-3.5ex} \phi_{\rm pulse}(\tau/\tau_0)\cos(\omega\tau) d\tau
=
\tau_0 \tilde{\psi}_{\rm pulse}(\omega\tau_0).
\end{equation}
Here, dependence on $\tau_0$ and $\omega\tau_0$ is from an integration against $\tau / \tau_0$.

To obtain the spectral intensity $I_j(\omega)$ for all of those pulses, their number density at each value of the width $\tau_0$ is necessary. Since the dimension of this density is~{\sf 1/T} while only $\tau_0$ is an available timescale, it is written as $n_{\rm c}/\tau_0$ with some dimensionless constant $n_{\rm c}$ \cite{mmym17}. We multiply Eq.~(\ref{eq17b}) by $n_{\rm c}/\tau_0$ and $j_{\rm c}^2$, integrate the result from $\tau_0 = \tau_{\rm small}$ to $\tau_{\rm large}$, and take the limits $\tau_{\rm small} \rightarrow 0$ and $\tau_{\rm large}  \rightarrow +\infty$,
\begin{align} \label{eq17c}
I_j(\omega) 
=           &\ n_{\rm c} j_{\rm c}^2\!        \int^{\tau_{\rm large}}_{\tau_{\rm small}}\! \tilde{\psi}_{\rm pulse}(\omega\tau_0)d\tau_0  \nonumber \\
\rightarrow    
            &\ n_{\rm c} j_{\rm c}^2\!        \int^{+\infty}         _0         \!\! \tilde{\psi}_{\rm pulse}(\omega\tau_0)d\tau_0       
\varpropto 
            \frac{j_{\rm c}^2}{\omega} .
\end{align}
\end{subequations}
The second integration has been done against $\omega\tau_0$ and has led to the scaling $\varpropto 1/\omega = 1/2\pi f$.

\begin{figure}[tbp]
\begin{center}
\resizebox{8.55cm}{!}{\includegraphics*[3.9cm,20.5cm][17.0cm,26.5cm]{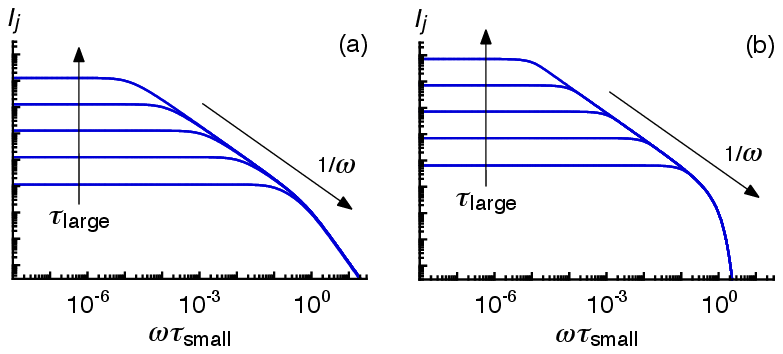}}
\caption{\label{f4} (a) Spectral intensity $I_j$ for $\phi_{\rm pulse} \varpropto \exp(-\tau/\tau_0)$ in Eq. (\ref{eq17}) at $\tau_{\rm large}/\tau_{\rm small} = 10^1$, $10^2$, ..., or $10^5$ against~the~normalized frequency $\omega \tau_{\rm small}$. (b) Same as in (a) but for $\phi_{\rm pulse} \varpropto \exp(-\tau^2/\tau_0^2)$.}
\end{center}
\end{figure}

Figure~\ref{f4} shows the cases for $\phi_{\rm pulse}(\tau/\tau_0) \varpropto \exp(-\tau/\tau_0)$ and $\varpropto \exp(-\tau^2/\tau_0^2)$. With an increase in the separation of $\tau_{\rm large}$ from $\tau_{\rm small}$, both of the two tend to scale as $1/\omega = 1/2\pi f$. Thus, under the condition of~Sec.~\ref{S4a}, any random superposition of random pulses is to reproduce the $1/f$ noise. At the lower frequencies $\omega \ll 1/\tau_{\rm large}$, $I_j(\omega)$ is white \cite{sv74, ngmprd81}. It is distinct from $I_{j_0}(\omega ) \varpropto \omega$ at $\omega \ll 1/\tau_{\rm c}$ in Eq.~(\ref{eq11a}) and Fig.~\ref{f1}(b).

\section{Concluding Remarks}  \label{S5}

For the occurrence of any stationary $1/f$ noise, we have formulated a necessary and sufficient condition. That is, if $j(t)$ is a random process with a mean amplitude $j_{\rm c} = [ \int^{\infty}_0\! I_j(\omega) d \omega ]^{1/2}$, its fluctuations need to be characterized by two constant timescales of $\tau_{\rm small} \ll \tau_{\rm large}$. At angular frequencies from $\omega = \omega_{\rm low} = 2\pi f_{\rm low} = 1/\tau_{\rm large}$~to~$\omega_{\rm high} = 2\pi f_{\rm high} = 1/\tau_{\rm small}$, we also need the absence of any other constant parameter. Then, in a range of $1/\tau_{\rm large} \ll \omega \ll 1/\tau_{\rm small}$, the spectral intensity $I_j(\omega)$ is obtained asymp-totically as $\varpropto j_{\rm c}^2/\omega$ or as $\varpropto j_{\rm c}^2/f$ [Eq.~(\ref{eq14})].

This condition is statistical, simple, and hence explains ubiquity observed for $1/f$ noise \cite{j25, p78, dh81, w88, pgfa14, sv74, mh76, mw69, thwb74, tsiggas90, mg86, b06, ngmprd81, lwpe87}. It is yet rigorous enough to explain why $1/f$ noise occurs in one system but not in very similar others \cite{w88, mw69}. Thus, it would serve as a statistically universal framework to discuss the physics of $1/f$ noise in each of the systems, e.g., about the wide separation between $\tau_{\rm small}$ and $\tau_{\rm large}$ \cite{ms82, btw87}.~The origin of $j_{\rm c}$ is also of interest, although its value is irrespective of the shape of $I_j(\omega)$. We ourselves have used the present condition to obtain Eq.~(\ref{eq16}), i.e., a generalization of the random pulse model of $1/f$ noise \cite{b37, h68}.

We have confirmed our condition with an example of $1/f$ noise found analytically for the flux $j_0(t)$ of the diffusion equation of Eq.~(\ref{eq5}) at the stationary boundary~of~Eq. (\ref{eq7}). For this noise [Eq.~(\ref{eq11})], a variety of unknown cases is likely (see Appendix for the equivalent electrical circuit). Conversely, if $j_0(t)$ were set to fluctuate with white noise, the diffusing quantity $q_0(t)$ at the boundary would exhibit $1/f$ noise as a result of Eq.~(\ref{eq10}).

The present study has followed those of scaling in fluids \cite{r26, l32, k41, ll59, b59, k62, t76, pa77, l78, agha84, sd98, lx10, mmym17, mm19, isy17}. As a final remark, we note intermittency of turbulence there \cite{k62, agha84, sd98, isy17}. This spatial or temporal inhomogeneity of intense fluctuations, albeit uncertain in detail, is known to change the power-law exponents anomalously from those derived in analytical studies analogous to ours. For example, from the Kolmogorov law $\varpropto k^{-5/3}$~of the velocity spectrum \cite{k41}, where turbulence is assumed to be not intermittent, the actual exponent differs by $0.04$--$0.05$ \cite{agha84, sd98, isy17}. A larger difference would be~observed if the turbulence were more intermittent.

While we have focused on noise of $\alpha = 1.0$ for~$1/f^{\alpha}$ \cite{j25, p78, dh81, w88, pgfa14, sv74, mh76, mw69, thwb74, tsiggas90, mg86, b06, ngmprd81, lwpe87}, the convincing cases of $\alpha \ne 1.0$ are also known \cite{dh81, w88}. They need to be studied separately. Nevertheless, as for a small difference from $\alpha = 1.0$, say, $\lesssim 0.1$, it might be due to the intermittency. Even when $\alpha = 1.0$ is expected for an intermittent case \cite{m80, krga06} or for a non-intermittent case \cite{b37, ba66, h68, ms82, btw87, mn23}, we might observe $\alpha \ne 1.0$ in the actual noise if its intermittency is somewhat different. To confirm this, higher-order statistics are necessary \cite{k62, agha84, sd98, isy17}. If the original exponent is indeed $\alpha = 1.0$, since~our~statistical condition has been formulated independently of any intermittency [Eq.~(\ref{eq15})], it holds at least as an approximation.

\vspace{-3mm}

\begin{acknowledgments} 
The author thanks M. Morikawa for useful discussions.
\end{acknowledgments}

\section*{Appendix}

\begin{figure}[tbp]
\begin{center}
\rotatebox{90}{
\resizebox{3.9cm}{!}{\includegraphics*[5.6cm,1.2cm][18.1cm,28.5cm]{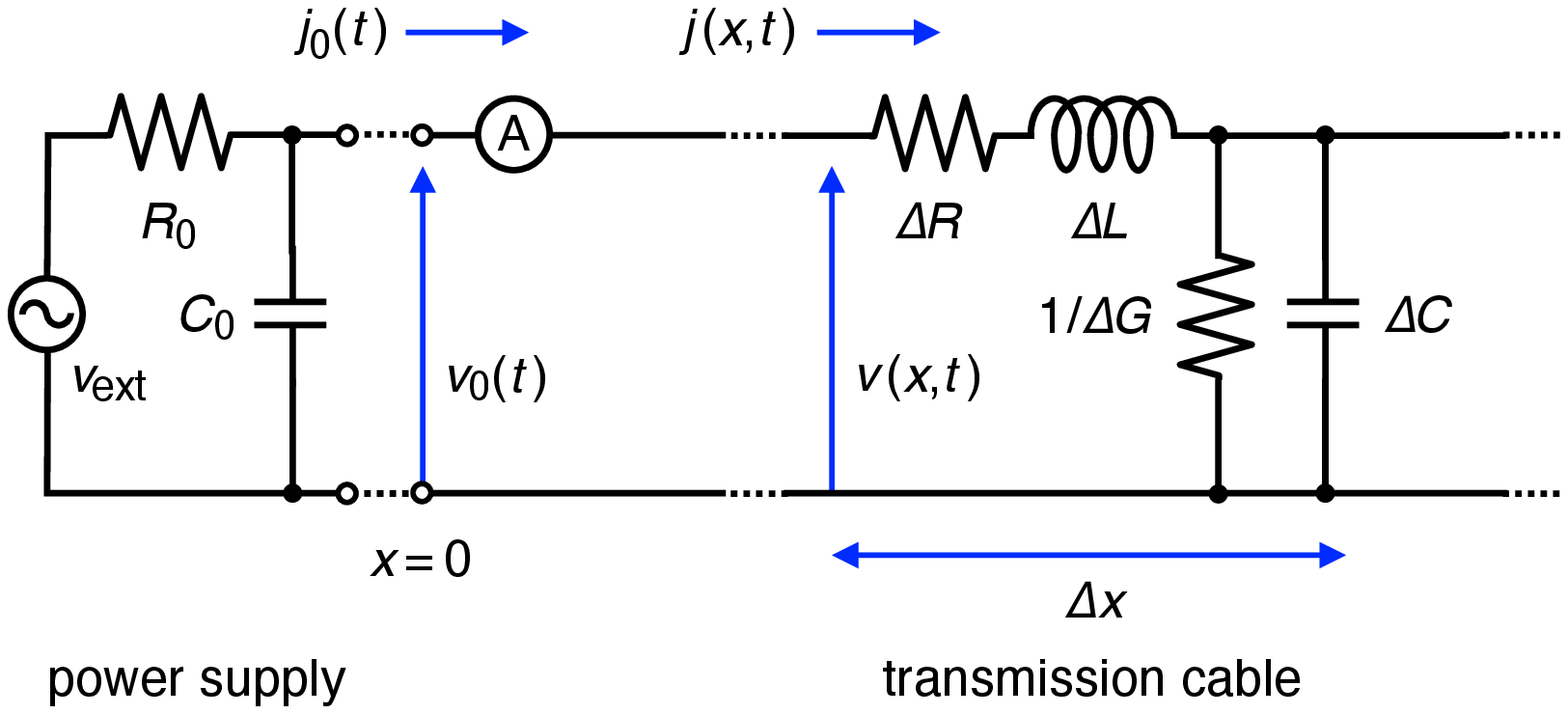}}
}
\caption{\label{f5} Electrical circuit made up of a transmission cable in Eq.~(\ref{eq18}) and a power supply in Eq.~(\ref{eq21}). The conductance~${\mit\Delta G}$ is just the inverse of the corresponding resistance.}
\end{center}
\end{figure}

Figure~\ref{f5} shows a practical example of a source of $1/f$ noise, i.e., an electrical circuit made up of a transmission cable and a power supply. Its behavior is the same as that of the diffusion equation of Eq.~(\ref{eq5}) under the boundary condition of Eq.~(\ref{eq7}).

The transmission cable is described by the telegraph equation \cite{ch62}. For the voltage $v$ and the current $j$ at any position $x$ from the terminal $x = 0$,
\begin{subequations} \label{eq18}
\begin{equation} \label{eq18a}
\frac{\partial}{\partial x}v(x, t) = -\frac{\mit\Delta R}{\mit\Delta x}j(x, t) -\frac{\mit\Delta L}{\mit\Delta x}\frac{\partial}{\partial t}j(x, t),
\end{equation}
and
\begin{equation} \label{eq18b}
\frac{\partial}{\partial x}j(x, t) = -\frac{\mit\Delta G}{\mit\Delta x}v(x, t) -\frac{\mit\Delta C}{\mit\Delta x}\frac{\partial}{\partial t}v(x, t).
\end{equation}
\end{subequations}
Here $\mit\Delta R$, $\mit\Delta L$, $\mit\Delta G$, and $\mit\Delta C$ denote the resistance, inductance, shunt conductance, and shunt capacitance of an infinitesimal segment $\mit\Delta x$ of the cable. Those per unit~cable length such as ${\mit\Delta R}/{\mit\Delta x}$ are constants.

By applying the Fourier transformation of Eq.~(\ref{eq1a}), we have
\begin{subequations} \label{eq19}
\begin{equation} \label{eq19a}
\frac{\partial}{\partial x} \tilde{v}(x, \omega) = -\frac{\mit\Delta R + i\omega \mit\Delta L}{\mit\Delta x}\, \tilde{j}(x, \omega) ,
\end{equation}
and
\begin{equation} \label{eq19b}
\frac{\partial}{\partial x} \tilde{j}(x, \omega) = -\frac{\mit\Delta G + i\omega \mit\Delta C}{\mit\Delta x}\, \tilde{v}(x, \omega) .
\end{equation}
\end{subequations}
Since $\mit\Delta G$ corresponds to the current leakage, it is usually negligible. We could hence set a wide range of the angular frequency ${\mit\Delta G}/{\mit\Delta C} \ll \omega \ll {\mit\Delta R}/{\mit\Delta L}$ to consider $\mit\Delta C$ and $\mit\Delta R$ alone. Then, Eq.~(\ref{eq18}) takes the form of Eq.~(\ref{eq5}),
\begin{subequations} \label{eq20}
\begin{equation} \label{eq20a}
\frac{\partial}{\partial t}v(x, t) = \frac{\mit\Delta x}{\mit\Delta C} \frac{\mit\Delta x}{\mit\Delta R} \frac{\partial^2}{\partial x^2}v(x, t) ,
\end{equation}
with
\begin{equation} \label{eq20b}
j(x, t) = -\frac{\mit\Delta x}{\mit\Delta R} \frac{\partial}{\partial x}v(x, t) .
\end{equation}
\end{subequations}
Since ${\mit\Delta C}{\mit\Delta R}$ has a dimension {\sf T}, $({\mit\Delta} x/{\mit\Delta} C) ({\mit\Delta} x/{\mit\Delta} R)$ has a dimension {\sf L}$^{\sf 2}${\sf /T} as in the case of the coefficient $D$ in~Eq. (\ref{eq5}).

The power supply has an external source $v_{\rm ext}$. Its output is set to fluctuate with the white noise $w(t)$, such~that the voltage $v_0(t)$ at the terminal $x = 0$ is described by~the Ornstein-Uhlenbeck process,
\begin{equation} \label{eq21}
\frac{dv_0(t)}{dt} = -\frac{v_0(t)}{C_0 R_0} + \sqrt{ \frac{4 \langle v_0^2 \rangle}{C_0 R_0}} w(t) .
\vspace{1mm}
\end{equation}
The corresponding spectrum is Lorentzian,
\begin{equation} \label{eq22}
I_{v_0}(\omega) = \frac{4C_0 R_0}{1+(C_0 R_0 \omega)^2} \langle v_0^2 \rangle.
\end{equation}
We have derived Eq.~(\ref{eq22}) from Eq.~(\ref{eq21}) in the same manner as for Eq.~(\ref{eq7b}) from Eq.~(\ref{eq12b}).

Through $D = ({\mit\Delta} x/{\mit\Delta} C) ({\mit\Delta} x/{\mit\Delta} R)$ and $\tau_{\rm c} = C_0 R_0$, Eqs.~(\ref{eq20}) and (\ref{eq22}) are equivalent to Eqs.~(\ref{eq5}) and (\ref{eq7}). The exception is the cutoff at $\omega = \omega_{\rm high} = {\mit\Delta R}/{\mit\Delta L}$. If ${\mit\Delta L}$ is so small that ${\mit\Delta R}/{\mit\Delta L}$ lies far above $\omega_{\rm low} = 1/C_0 R_0$, the current $j_0(t)$ at the terminal $x = 0$ exhibits the noise of $1/f = 2\pi / \omega$ in a range of $1/C_0 R_0 \ll \omega \ll {\mit\Delta R}/{\mit\Delta L}$.~At the higher frequencies $\omega \gg {\mit\Delta R}/{\mit\Delta L}$, we could ignore ${\mit\Delta R}$ as well as ${\mit\Delta G}$ in Eq.~(\ref{eq19}) to obtain a wave equation and hence $I_{j_0}(\omega) \varpropto I_{v_0}(\omega) \varpropto 1/\omega^2$.

\end{document}